\documentclass[a4paper,11pt,authoryear]{elsarticle}
\usepackage{pdfpages}

\usepackage{import, graphicx, xcolor, calc}
\usepackage{subfigure}
\usepackage{hyperref}
\usepackage{float}
\usepackage[modulo]{lineno}

\usepackage[fleqn]{amsmath}
\usepackage{bbold}
\usepackage{physics}
\usepackage{mathtools}

\journal{Journal}

\floatstyle{plaintop}
\restylefloat{table}

\usepackage[ left=2cm, right=2cm, top=2.5cm, bottom=3cm, includehead]{geometry}

\newcommand{\mbf}[1]{\mathbf{#1}}

\begin{document}

\begin{frontmatter}
    \title{Multiscale Cavitation Sub-Grid Modeling via Population Balances as Linear Stochastic Process}

    \author{F. J. Aschmoneit\corref{mycorrespondingauthor}}
    \ead{fynnja@math.aau.dk}
    \address{Department of Mathematical Sciences, Aalborg University, Copenhagen, Denmark}
    
\linenumbers

\begin{abstract}
    A multiscale sub-grid cavitation model is developed in which the bubble size distribution evolves as a linear stochastic process in radius space.
    Starting from the integrated Rayleigh--Plesset equation, the population balance is recast as a hyperbolic transport equation for the number density per radius, whose method-of-characteristics solution, projected onto a discrete histogram basis, yields a column-stochastic Markov chain governing the bubble counts per size bin. 
    The transition matrix factors into a precomputable, mesh-only geometric part and a local, pressure-dependent shift, isolating the coupling to the surrounding flow into a single dimensionless vector per cell. The framework recovers classical homogeneous-mixture cavitation closures in the limit of a single representative scale.
\end{abstract}


\end{frontmatter}

\section{Introduction}

    Dispersed cavities are tracked over multiple size scales, ranging from the smallest inception scale to a macroscopic scale where the assumption of spherical cavities no longer holds.
    The cavities grow and shrink as a consequence of competing effects of vapor pressure and surface tension.
    While the vapor pressure is constant for all cavities, the opposing surface tension depends on the specific cavity scale.
    The rate of bubble growth or shrinking therefore differs across scales.
    Therefore, the number density of bubble size scales is tracked across the different scales, employing integrated Raleigh-Plesset equation (RPE).
    
    In the following I denote a vector with in bold face $\mbf{a}$ and its elements as $a_i$.
    The inner product is denoted as $\mbf{a}\cdot \mbf{b}$.
    A matrix is denoted with a capital letter $\Omega$, while its elements are denoted as $\Omega_{ij}$. 
    The matrix-vector product is either denoted as $\Omega \mbf{a}$.
    The element-wise multiplication of two vectors is $\mbf{a} \circ \mbf{b}$, and taking the element-wise cube of a vector uses $\mbf{a}^{\circ 3}$.

\section{Continuous two-phase flow with cavitation}
    The conservation of mass and momentum are governed by the continuity equation and the Navier-Stokes equations below.
    
    \begin{equation}\label{eq:continuityEq}
        \frac{\partial}{\partial t}\rho + \div (\rho \mbf{u}) = 0
    \end{equation}

    \begin{equation}\label{eq:NSE}
        \frac{\partial}{\partial t} \mbf{\rho u} + \big( \mbf{u} \cdot \grad \big)  \rho \mbf{u}  = -\grad p +  \div( \mu  \grad \mbf{u}) 
    \end{equation}

    In the homogeneous-mixture approach, fluid density $\rho$ and viscosity $\mu$ are linear mixtures of the vapor fraction $\alpha$.
    In this study, $\alpha$ is a derived quantity, see Eq.~\ref{eq:alphaDef}, and hence, it does not follow a transport equation, as in typical Volume-of-Fluid (VoF) cavitation methods, see also \cite{folden_classification_2023}.
    
    \begin{align}
    \rho &= \alpha \rho_V + (1-\alpha)\rho_L \label{eq:RhoMuMixture}\\
    \mu &= \alpha \mu_V + (1-\alpha)\mu_L
    \end{align}

    In the homogeneous-mixture framework, Eq.~\ref{eq:RhoMuMixture} serves as an equation of state, relating the mixture density to the vapor fraction $\alpha$, while Eq.~\ref{eq:continuityEq} closes the pressure--velocity coupling. 
    Standard simplifications such as the weakly-compressible assumption, together with established discretizations (e.g. FVM with pressure--velocity coupling schemes such as SIMPLE or PISO, or lattice Boltzmann methods, FEM, ...), can be applied. 
    The focus of this article is the population-balance cavitation model itself, independently of any particular CFD implementation.
    Note that due to the non-constant density, that $\vb{u}$ is not divergence-free, see Eq.~\ref{eq:velocityDivergence}.

\section{Sub-grid Bubble Dynamics}
    I am assuming that the cavity pressure is constant at $p_{sat}$ and the liquid pressure changes at a much slower rate than the bubble size, and hence $dr/dt \approx \dot r$.
    Phase transition is modeled as the growing/shrinking of spherical bubbles, according to RPE in Eq.~\ref{eq:RPE}, where $r$ denotes the bubble radius.
    $p$ is the ambient liquid static pressure from Eq.~\ref{eq:NSE}, which is assumed to be in quasi-equilibrium during bubble expansion/contraction. 
    The non-condensable gas pressure $p_{ge} \tfrac{r_e^3}{r^3}$ is anti-proportional to the bubble volume.

    \begin{equation}\label{eq:RPE}
        \rho_L \qty( r \Ddot r + \frac{3}{2} \dot r^2 ) = p_{ge} \tfrac{r_e^3}{r^3} + p_{sat} - p - \frac{4 \mu_L}{r} \dot{r} - \frac{2 \sigma}{r}
    \end{equation}

    Neglecting the viscous term, and identifying $r \Ddot r + \frac{3}{2} \dot r^2 = \frac{1}{2\dot r r^2} \dv{t}\qty( \dot r^2 r^3)$, the PRE is integrated from the nuclei equilibrium radius $r_{e}$ up to $r$, see \cite{franc_fundamentals_2010}.
    \begin{equation}\label{eq:integralRPE} 
        \rho \, \dot r^2 = \frac{3}{2} (p_{sat} - p) \left( 1 - \frac{r_e^3}{r^3}  \right) 
        + 2 p_{ge} \frac{r_e^3}{r^3} \ln{\left(\frac{r}{r_e}\right)} 
        - \frac{2\sigma}{r} \left(  1 - \frac{r_e^2}{r^2} \right)
    \end{equation}

\begin{figure}[H]
        \centering
        \includegraphics[width=0.49\linewidth]{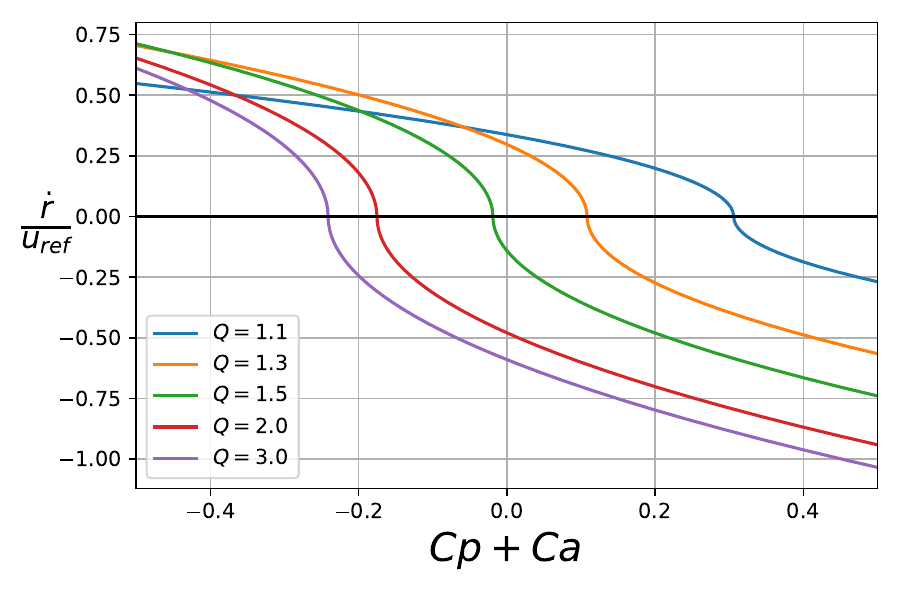}
        \hfill
        \includegraphics[width=0.49\linewidth]{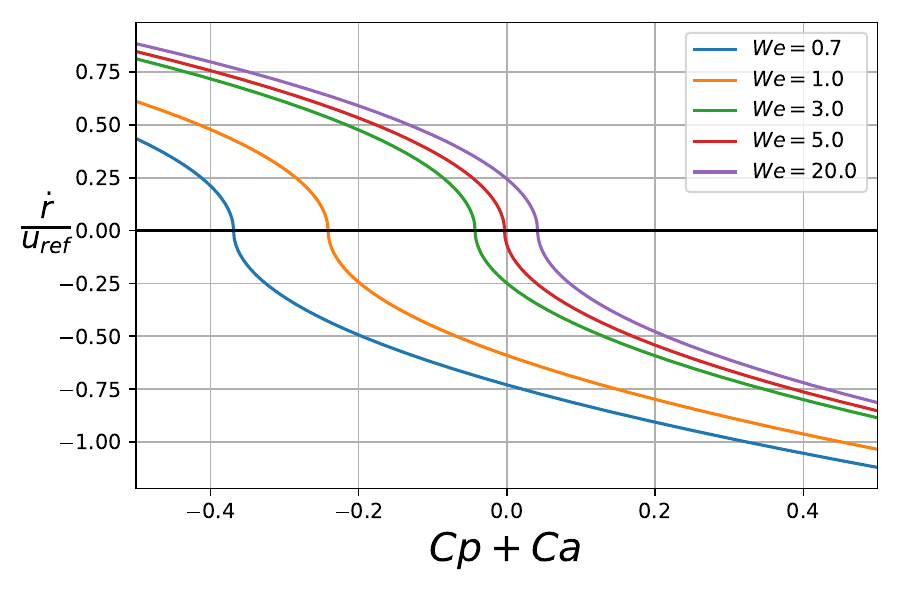}
        \caption{Bubble growth/shrinking according to Eq.~\ref{eq:dotR_nonDimensional}, left: We = 1, right: Q = 3}
        \label{fig:dotR_nonDimensional}
    \end{figure}

    With $p_{ge}=p_\infty - p_v + \tfrac{2 \sigma}{r_e}$. 
    Here, I define the reference velocity as the bubble expansion velocity $u= \sqrt{\tfrac{p_\infty - p_v}{\rho}}$.
    In non-dimensional terms the cavitating flow is characterized by the cavitation number $Ca = \frac{p_\infty - p_v}{\frac{1}{2} \rho u^2}$ and the pressure coefficient $Cp = \frac{p - p_\infty}{\frac{1}{2} \rho u^2}$.
    With 
    $P=Cp+Ca = \tfrac{p - p_v}{p_\infty - p_v}$,
    $Q=\tfrac{r}{r_e}$, 
    $We = \tfrac{\rho \, r_e \, u^2}{\sigma}=\tfrac{r_e p_{ge}}{\sigma} - 2$, 
    the RPE is written in non-dimensional form below. 
    Fig. \ref{fig:dotR_nonDimensional} illustrates the effect of different bubble radii and Weber numbers on the growth rate, stressing the necessity for a multi-scale approach.
    \begin{equation}\label{eq:dotR_nonDimensional}
        \qty( \frac{\dot r}{u} )^2
        =   -\frac{3}{2} P \qty( 1 - Q^{-3})  
            + 2\qty(1 + \frac{2}{We})Q^{-3}\ln{\qty(Q)} 
            - \frac{2}{We} \qty( Q^{-1} - Q^{-3} ) 
    \end{equation}
    
    The dispersion of cavity bubbles in the sub-grid are depicted through the number density per radius $f(t,r)$, with $\qty[f]=m^{-4}$.
    $f(r,t)$ evolves under Rayleigh--Plesset growth as a pure advection in radius space, where $\dot r$ is the local growth rate from the integral RPE, Eq.~\ref{eq:integralRPE}.
    
    \begin{equation}\label{eq:rspaceAdvection}
        \pdv{f}{t} + \pdv{r}\bigl( \dot r\,f \bigr) \;=\; 0
    \end{equation}
    
    With knowing the evolution of $f$ the evolution of the different moments link to the how the microscopic cavities affect the macroscopic flow.
    Its $k-$th moment is defined as:
    
    \begin{equation}\label{eq:distrMomentsContinuous}
        \mu_{k} = \int_{r_{min}}^{r_{max}} r^k f(t,r) \dd{r}
    \end{equation}

    The interval of spherical bubble sizes considered is $\mathcal{R} = \qty[0, R_{max}]$ is dissected into $M$ sub-intervals $\mathcal{R}_i = \qty[R_{i-1}, R_i]$ of width $w_i = R_i - R_{i-1}$, with $i=1,2,...,M$.
    Introducing the piece-wise constant $\bar{f}_i$ over $\mathcal{R}_i$, $f(t,r)$ is approximated to order $\order{ \tfrac{1}{M}}$.
    Assuming $f(t,r)$ to be integrable across $\mathcal{R}$, $\mu_k$ in Eq.~\ref{eq:distrMomentsContinuous} may be splitted into a series over   $\mathcal{R}_i$, with the distribution moments per bin $\mu_{k,i}$. 

    \begin{equation}\label{eq:sumOfMoments}
        \mu_k \approx \sum_{i=1}^M \underbrace{\bar{f_i} \frac{R^{k+1}_i - R^{k+1}_{i-1}}{k+1} }_{\mu_{k,i}}
    \end{equation}
    
    where $\mu_{0,i} = \bar{f_i}  w_i := n_i$ is the number density of nuclei or bubbles with radii uniformly distributed over $\mathcal{R}_i$, 
    and $\mu_{3,i}  =  n_i \frac{R_{i-1}^3 + R_i^3}{2} + \order{\Delta R_i^2} \approx n_i r_i^3$ is proportional to the volume fraction of cavities with representative size $r_i = \sqrt[3]{\tfrac{R_{i-1}^3 + R_{i}^3}{2}}$.
    In vector notation, the $\alpha$ follows as a simple scalar product of $\vb{n}$ and $\vb{r}^{\circ 3}$ (element-wise power) as:
    
    \begin{equation} \label{eq:alphaDef}
        \alpha = \frac{4}{3} \pi \mu_{3}    = \frac{4}{3} \pi \, \mbf{n} \cdot \mbf{r}^{\circ 3}
    \end{equation}

    So how do the $n_i$ evolve in time?
    Eq.~\ref{eq:rspaceAdvection} is a continuity equation in radius space developing $f$.
    By the method of characteristics, $f$ is conserved along characteristic curves:
    
    \begin{equation}\label{eq:charCurves}
        \int_{\mathcal{R}_i} f(r, t + \delta t) dr = \int_{\mathcal{R}_i-\dot r (r^\prime) \delta t} f(r^\prime, \delta t) dr
    \end{equation}

    Substituting the piecewise-constant $\bar f_i$, the left-hand side becomes $n_i(t+\delta t)$ and the right-hand side reduces to a sum of integral-overlaps with bin $\mathcal{R}_i$, where the overlaps are a consequence of a share of bubbles in $\mathcal{R}_j$ grow/shrink into $\mathcal{R}_i$.
    If these overlap integrals are collected in the transition matrix $\Gamma_{ij}$, Eq.~\ref{eq:charCurves} reduces to a linear update in Eq.~\ref{eq:nMarkov_index} or a  vector-matrix product in Eq.~\ref{eq:nMarkov_matrix}.

     \begin{align}
        &n_i(t+\delta t) \;=\; \sum_{j=1}^{M} \Gamma_{ij}\, n_j(t) \label{eq:nMarkov_index} \\
        &\mbf{n}_{t+\delta t} = \Gamma \mbf{n}_{t} \label{eq:nMarkov_matrix}
    \end{align}
    
    The elements in $\Gamma$ are defined as overlap of the shifted interval $\mathcal{R}_j +  \dot r_j \delta t$ with the target interval $\mathcal{R}_i$, relative to width of the source interval $ \abs{\mathcal{R}_j }$:

    \begin{equation}\label{eq:gammaDef1}
        \Gamma_{ij} = \frac{\qty| \qty( \mathcal{R}_j + \dot r_j \delta t ) \cap \mathcal{R}_i |}{\abs{\mathcal{R}_j }}    
        \qquad i,j \in \{1,\dots,M\}
    \end{equation}

    The bounding target intervals $\mathcal{R}_1, \mathcal{R}_M$ need special treatment, in order to capture potential over/under shooting. 
    If $i=1: \mathcal{R}_1 \longrightarrow \mathcal{R}_1 \cup \qty(-\infty, R_{min})$ and if $i=M: \mathcal{R}_M \longrightarrow \mathcal{R}_M \cup \qty(R_{max}, \infty)$.
    The lower-triangular entries, are evaporation coefficients, while the upper-triangular are condensation coefficients.
    From $\Gamma_{ij}$, with $0\leq \Gamma_{ij} \leq 1$ and $\sum_i \Gamma_{ij´} = 1$ follows that $\Gamma$ is a column-stochastic matrix and Eq.~\ref{eq:nMarkov_matrix} is a Markov process.

    Since $\dot r$ depends on the local pressure, $\Gamma$ must be evaluated locally, requiring an efficient method.
    Defining the global, constant distance matrix of boundaries in Eq.~\ref{eq:distMat}. 
    It follows $D_{ij} = -D_{ji}$, $D_{ij} = D_{ik} - D_{jk}$, and $\abs{\mathcal{R}_j } = D_{j j-1}$.
    \begin{equation}\label{eq:distMat}
        D_{ij} = r_i - r_j \qquad i,j \in \{0,\dots,M\}
    \end{equation} 

    The local shifts in bubble sizes are expressed in non-dimensional form through the vector of Courant numbers $\mathbf{C}$.
    Due to its dependence on $\delta p$, $\mathbf{C}$ is evaluated locally, and may change at every time step.
    \begin{equation}
        C_j = \frac{\dot R_j \delta t}{ D_{jj-1} } \qquad j \in \{1,\dots,M\}
    \end{equation}
    
    The transition matrix $\Gamma$ is assembled as a function of the global $D$ and the local $\mathbf{C}$.
    With $g(x)=max(0,x)$ denoting the ramp or Relu function, Eq.~\ref{eq:gammaDef1} is written in a practical, functional form below.
    In order to avoid over- and undershooting, it is sufficient to discard term 1 if $i=M$, or term 2 if $i=1$.
    
    \begin{equation}
        \Gamma_{ij} = g \qty( 1  
        - \underbrace{g \qty( C_j - \frac{D_{ij}}{D_{jj-1}} ) }_1
        - \underbrace{g \qty( \frac{D_{i-1j-1}}{D_{jj-1} }  - C_j ) }_2 )
        \qquad i,j \in \{1,\dots,M\}
    \end{equation}

    The velocity divergence follows from Eq.~\ref{eq:continuityEq}. subject to Eqs.~\ref{eq:RhoMuMixture},~\ref{eq:alphaDef} as:
    
    \begin{equation}\label{eq:velocityDivergence}
        \div \mbf{u} = \frac{4}{3} \pi \left(  1 + \frac{\rho_V}{\rho_L}  \right) \vb{\dot{n}}\cdot{\mbf{R}^{\circ 3}}
    \end{equation}

    With
    
    \begin{equation}
         \mbf{\dot{n}} 
        = \frac{\mbf{n}_{t+ \delta t} - \mbf{n}_t}{ \delta t} = \frac{( \Gamma -  \mathbb{1}) \mbf{n} }{\delta t} \label{eq:numberDensitySource}
    \end{equation}

\section{Sub-Grid Discretization}
    
    The volume difference across the bubble interval grows with a factor $\lambda\geq 1$: $R_{i+1}^3 - R_i^3 = \lambda ( R_i^3 - R_{i-1}^3 )$, see also \cite{liao_discrete_2018}.
    It is practical to adjust the Courant number with $\lambda$, after $R_{min},R_{max}, M$ and $\delta t$ have been set.
    The linear system (\ref{eq:boundaryDiscretization}) evaluates the cube of the bubble interval endpoints. 
    Omitted entries in sparse matrices are zero.
    
    \begin{equation}\label{eq:boundaryDiscretization}
        \mqty( 
        \dmat{
        1&&&\\
        -\lambda&1+\lambda&-1&\\
        &-\lambda&1+\lambda&-1\\
        &&&\ddots,
        -\lambda&1+\lambda&-1
        \\&&1}   
        ) \cdot
        \mqty(R_0^3\\R_1^3\\  \vdots \\ \\ R_{M-1}^3\\ R_M^3) = 
        \mqty(R_{min}^3\\0\\ \vdots \\ \\ 0\\ R_{max}^3)
    \end{equation}
    
    When all $N$ bubbles have grown to the largest scale $R_{max}$, I assume that the entire unit volume is occupied with vapor, which couples the greatest bubble scale and the total number of bubbles through $N \frac{4}{3} \pi R_{max}^3 = 1$, so either of these variables may be treated as a dependent variable.

\section{Conclusions}
    A novel multi-scale cavitation model is presented.
    It's population balance approach allows to model the dynamics of cavitation bubbles in the sub-grid as a simple Markov Chain.
    At this stage, the model does not incorporate coalescence/break up effects.

\bibliography{references.bib}
\end{document}